\documentclass[a4paper,graphics,natbib,graphicx,psfig,amssymb,ccaption]{article}
\usepackage{lscape}
\usepackage{graphicx}
\newcommand{\affil}[1]{$^{\rm #1}$}
\usepackage[authoryear]{natbib}
\bibpunct{(}{)}{;}{a}{}{,}
\date{} 

\title{\large\bf\flushleft Transformations between WISE and 2MASS, SDSS, BVI photometric systems:
II. Transformation equations for red clump stars}
\author{\parbox{\textwidth}{\flushleft
\vspace{-0.5cm}
{\it S. Bilir\affil{A}, S. Karaali\affil{A}, N. D. Da\u gtekin\affil{A}, \"O. \"Onal\affil{A}, S. Ak\affil{A}, T. Ak\affil{A} and A. Cabrera-Lavers\affil{B, C, D}}\\
\vspace{0.4cm}
{\small \affil{A}\,Istanbul University, Faculty of Sciences, Department 
of Astronomy and Space Sciences, 34119 University, Istanbul, Turkey, Email: sbilir@istanbul.edu.tr}\\
{\small \affil{B}\,Instituto de Astrof\'{\i}sica de Canarias, E-38205 La Laguna, Tenerife, Spain}\\
{\small \affil{C}\,GTC Project Office, E-38205 La Laguna, Tenerife, Spain}\\
{\small \affil{D}\,Departamento de Astrof\'{\i}sica, Universidad de La Laguna, E-38205 La Laguna, Tenerife, Spain}\\
}}
\begin{document}
\twocolumn[
\begin{changemargin}{.8cm}{.5cm}
\begin{minipage}{.9\textwidth}
\vspace{-1cm}
\maketitle
%
%
\small{{\bf Abstract:} We present colour transformations for the conversion of the {\it Wide-Field Survey Explorer (WISE)} $W1$, $W2$, and $W3$ magnitudes to the Johnson-Cousins ($BVI_c$), Sloan Digital Sky Survey ($gri$), and Two Micron All Sky Survey $JHK_s$ photometric systems, for red clump (RC) stars. RC stars were selected from the Third Radial Velocity Experiment (RAVE) Data Release (DR3). The apparent magnitudes were collected by matching the coordinates of this sample with different photometric catalogues. The final sample (355 RC stars) used to obtain metallicity dependent-and free of metallicity- transformations between {\em WISE} and Johnson-Cousins, SDSS, 2MASS photometric systems. These transformations combined with known absolute magnitudes at shorter wavelengths can be used in space density determinations for the Galactic (thin and thick) discs at distances larger than the ones evaluated with $JHK_{s}$ photometry alone, hence providing a powerful tool in the analysis of Galactic structure.        
}

\medskip
\medskip
\end{minipage}
\end{changemargin}
]
\small

\section{Introduction}

Red clump (RC) stars are core helium-burning stars, in an identical evolutionary phase to those stars which make up the horizontal branch in globular clusters. However, in intermediate -and higher- metallicity systems only the red end of the distribution is seen, forming a clump of stars in the colour-magnitude diagram. In recent years much work has been devoted to studying the suitability of RC stars as a distance indicator. Their absolute magnitude in the optical ranges from $M_V=+0.70$ mag for those of spectral type G8 III to $M_V=+1.0$ mag for the K2 III ones \citep{Keenan99}. The absolute magnitude of these stars in the $K_s$ band is $M_{K_s}-1.54\pm0.04$ mag with negligible dependence in metallicity \citep{Groenewegen08}. The optical and infrared colour ranges for these stars are $0.8\leq (B-V)_0\leq1.3$ and $0.29\leq(J-H)_0\leq0.65$, respectively, and they have a limited surface gravity, i.e. $2.1\leq \log g~(cms^{-2})\leq2.7$ \citep{Puzeras10}. It should be added that RC stars are different in structure than the ones in late transitional phases of evolution off main sequence or immediately before a supernove which have circumstellar material, such as red supergiants, yellow hypergiants, luminous blue variables, B[e] supergiants and equatorial rings, interacting binaries, and Wolf-Rayet stars.

In a former paper \citep[][hereafter Paper I]{Bilir11a} we presented the transformation equations between {\it WISE} and 2MASS, SDSS, Johnson-Cousins photometric systems for dwarf stars. Here, our aim is to obtain similar transformations between the same photometric systems but for RC stars. The galactic model parameters can be obtained more precisely using {\it WISE} absolute magnitudes calculated from these transformations. In the next paragraphs, we give a short definition for the mentioned photometric systems and the RAVE survey, which provides the data used in our study. However, we refer the reader to Paper I for a more complete information.

The Sloan Digital Sky Survey (SDSS) obtains images almost simultaneously in five broad bands ($u$, $g$, $r$, $i$, and $z$) centered at $3560$, $4680$, $6180$, $7500$, and $8870$ \AA, respectively, \citep{York00}. The magnitudes derived from fitting a point spread function (PSF) are currently accurate to about 1 per cent in $g$, $r$, $i$, $z$ and 2 per cent in $u$ for point sources \citep{Padmanabhan08}. The data have been made public in a series of yearly data release where the eighth data release \citep[DR8,][]{Aihara11} covers 14 555 deg$^{2}$ of imaging area. The limiting magnitudes are ($u$, $g$, $r$, $i$, $z$) = (22, 22.2, 22.2, 21.3, 20.5).

The Two Micron All Sky Survey \citep[2MASS;][]{Skrutskie06} provides the most complete data base of near infrared (NIR) Galactic point sources available to date. Observations cover 99.998 per cent \citep{Skrutskie06} of the sky with simultaneous detections in $J$ (1.25 $\mu$m), $H$ (1.65 $\mu$m), and $K_{s}$ (2.17 $\mu$m) bands up to the limiting magnitudes of 15.8, 15.1, and 14.3, respectively. Bright source extractions have 1$\sigma$ photometric uncertainty of $<$0.03 mag and astrometric accuracy on the order of 100 mas.

The Wide Field Survey Explorer \citep[{\it WISE};][]{Wright10}, an up-to-date infrared (IR) survey, began surveying the sky on 2010 January 14 and completed its first full coverage of the sky on 2010 July 17 with much higher sensitivity than comparable previous IR survey missions. {\it WISE} has four IR filters $W1$, $W2$, $W3$, and $W4$ centered at 3.4, 4.6, 12, and 22 $\mu$m, and with the angular resolutions 6.1, 6.4, 6.5, and 12 arcsec, respectively and has a 40-cm telescope feeding array with a total of four million pixels. {\it WISE} has achieved 5$\sigma$ point source sensitivities better than 0.08, 0.11, 1, and 6 mJy at 3.4, 4.6, 12, and 22 $\mu$m, respectively. These sensitivities correspond to the Vega magnitudes 16.5, 15.5, 11.2, and 7.9. Thus {\it WISE} will go a magnitude deeper than the 2MASS $K_{s}$ data in $W1$ for sources with spectra close to that of an A0 star, and even deeper for moderately red sources like K stars or galaxies with old stellar populations. 

The Radial Velocity Experiment \citep[RAVE,][]{Steinmetz06} measures radial velocities and stellar atmospheric parameters from spectra using the 6dF multi-object spectrometer on the Anglo-Australian  Astronomical Observatory's 1.2-m UK Schmidt Telescope. The survey looks in the Ca-triplet region (8410-8795\AA), has a resolution of $\sim$7500, and is magnitude limited. The targets chosen are southern hemisphere stars taken from the {\em Tycho-2}, SuperCOSMOS and the Deep Near Infrared Survey of the Southern Sky \citep[DENIS,][]{Fouque00} surveys with $I$-band magnitudes between 9 and 13. The average internal errors in radial velocity are $\sim$2 km s$^{-1}$, and the approximate radial velocity offset between the RAVE and the literature is smaller than $\sim$1 km s$^{-1}$. The catalogue also includes 2MASS photometry and proper motions from Starnet 2.0, {\em Tycho-2}, SuperCOSMOS, and UCAC2 \cite[for more information about RAVE, see][]{Zwitter08}.     

The passband profiles for the Johnson-Cousins, SDSS, 2MASS, and {\it WISE} photometric systems are shown in Fig. 1. With respect to the same figure in Paper I, we omitted here the passband for $R$ which is not used in our transformations but we added the DENIS passband $I_{d}$ with which we evaluated the Cousins optical magnitudes ($I_{c}$). $W3$ and $W4$ could not be used in Paper I in the transformations for dwarfs due to the faintness of dwarfs in both bands. In this study of RC stars, the $W3$ magnitudes could be used but the $W4$ magnitudes were too faint. Hence, as in the inverse transformations for dwarfs, the $J-H$ colour of the 2MASS photometric system is used as a second colour combined linearly with $W1-W2$. Fig. 2 plots the fields available with the {\it WISE} and RAVE surveys. 

\begin{figure}
\begin{center}
\includegraphics[scale=0.45, angle=0]{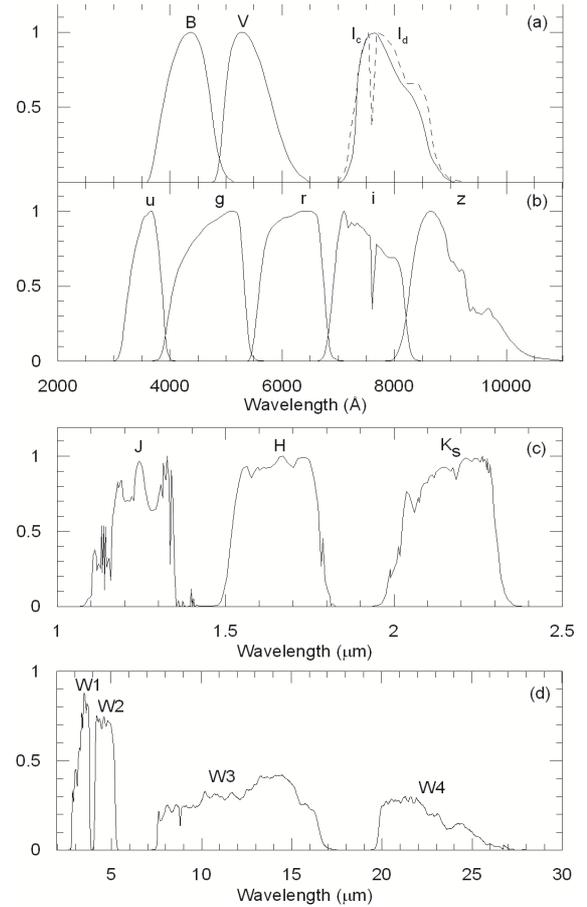}
\caption[] {Normalized passbands of the Johnson-Cousins-DENIS filters (a), the SDSS filters (b), the 2MASS filters (c), and the {\it WISE} filters (d).}
\end{center}
\end{figure}

\begin{figure}
\begin{center}
\includegraphics[scale=0.40, angle=0]{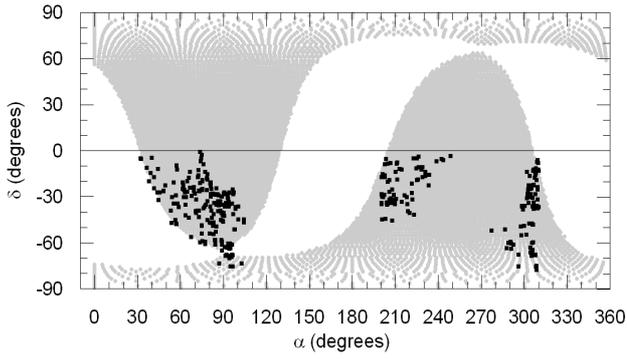}
\caption[] {Equatorial coordinates of the stars observed in {\it WISE} (grey regions) and RAVE (black squares) surveys.}
\end{center}
\end{figure}

In Section 2 we present the sources of our sample and the criteria applied to the chosen stars. The transformation equations are given in Section 3 and finally, we give a summary and conclusions in Section 4.

\section{The data}
\subsection{RAVE Sample with {\em Tycho-2}, DENIS, and 2MASS Data}
The main source of our data is the RAVE Data Release (DR3) catalogue \citep{Siebert11}. The reason of this choice is the advantage of the RAVE catalogue. This catalogue includes the atmospheric parameters ($T_{eff}$, $\log g$, $[M/H]$) with high accuracy. This is important, as the surface gravity is used to separate the dwarfs and the RC stars and the transformations are derived for different metallicity bins. We initially applied two constraints: $2<\log g~(cms^{-2})\leq 3$ and $J-H>0.4$, and obtained a sample of 8003 stars from the RAVE DR3. The reason of these constraints is due to the fact that most of the RC stars lie in this $\log g$ interval and that they are much larger in number in the $J-H>0.4$ colour interval \citep{Bilir11b}. We then included the following additional but necessary constraints: i) We selected stars for which {\em Tycho-2} ($B_{T}$, $V_{T}$), DENIS ($I_{d}$), and 2MASS ($JHK_s$) magnitudes were available (3103 stars), ii) we matched the reduced RAVE DR3 catalogue with the {\it WISE} Preliminary Data Release (PDR)\footnote{http://irsa.ipac.caltech.edu/cgi-bin/Gator/nph-scan?mission=irsa\&submit=Select\&projshort=WISE$\_PRELIM$} catalogue and chose the stars which were available with $W1$, $W2$, $W3$, and $W4$ magnitudes (954 stars), iii) we used the magnitude flags, labeled ``AAA'', which means $S/N\geq10$, i.e. they have the highest quality measurements, for the 2MASS and {\it WISE} magnitudes (918 stars), iv) we limited $B_{T}-V_{T}$ colours with $0.8<(B_{T}-V_{T})\leq1.7$ and excluded the stars with $B_{T}-V_{T}$ error larger than 0.2. Thus the complete sample reduced to 355 stars. The $(J-H)_0-(B-V)_0$ two colour diagram and the spectral distribution of the final sample in three metallicity categories are given in Fig. 3.

\subsection{Evaluation the $BVI_{c}$ magnitudes}
The RAVE survey does not involve any star observed in the Johnson-Cousins ($BVI_{c}$) system. Hence, we have to use the following procedure to obtain $B$, $V$, and $I_{c}$ magnitudes for our sample: We revealed that 370 stars in the Landolt's (2009) $UBVRI$ Photometric Standard Stars catalogue were observed in the DENIS survey \citep{Fouque00}. We excluded stars with errors in $I_d$ larger than 0.1 mag, thus the sample reduced to 355. We matched this sample with the 2MASS catalogue and used the magnitude flags, labeled ``AAA'', for obtaining the magnitudes of highest quality. This constraint reduced the number of stars to 344. 

Finally, we plotted the $V$ magnitudes of Johnson versus the $J$ magnitudes of 2MASS in a two magnitude diagram and eliminated the dwarfs from the sample in Fig. 4 \citep[see][for a description about the elimination method]{Bilir06}. By doing this, the final giant sample consists of 128 stars.

\begin{figure}
\begin{center}
\includegraphics[scale=0.40, angle=0]{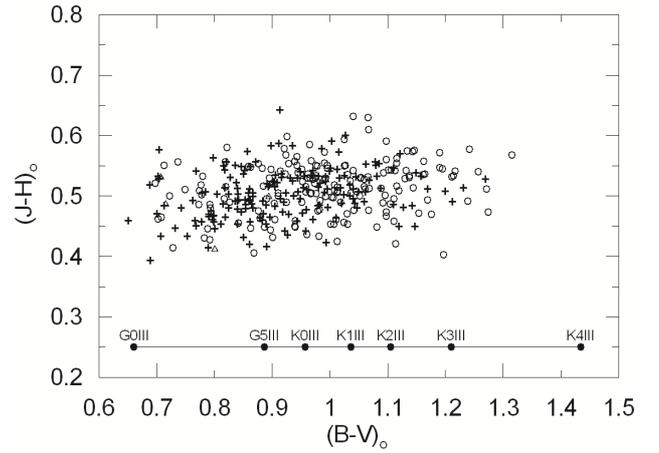}
\caption[]{Two colour diagram of the sample stars. The symbols give: ($\circ$) $[M/H]>-0.4$, ($+$) $-1<[M/H]\leq-0.4$, and ($\bigtriangleup$) $[M/H]\leq-1$ dex.}
\end{center}
\end{figure}
Fig. 4b compares the optical magnitudes of DENIS ($I_{d}$) and Cousins ($I_{c}$) supplied from \citet{Landolt09}. After rejecting four stars which showed large scattering, we obtained the following equation which is used for evaluation of the $I_{c}$ (hereafter $I$) magnitudes of the sample:

\begin{table*}
\setlength{\tabcolsep}{.1pt}
{\scriptsize
\center
\caption{Johnson-Cousins, SDSS, 2MASS and {\it WISE} magnitudes and colours of the sample stars (in total 355 stars). The columns give: (1) star name; (2)
and (3) Galactic coordinates; (4) $V_o$-apparent magnitude; (5) and (6) $(B-V)_o$ and $(V-I)_o$ colour indices, respectively; (7) $g_o$-apparent magnitude; (8) and (9) $(g-r)_o$ and $(r-i)_o$ colour indices, respectively; (10) $J_o$-apparent magnitude; (11) and (12) $(J-H)_o$ and $(H-K_s)_o$ colour indices; (13) $W1_o$-apparent magnitude; (14)  and (15) $(W1-W2)_o$ and $(W2-W3)_o$ colour indices; and (16) reduced $E_d(B-V)$ colour excess, respectively. The complete table is available in electronic format (see Supporting Information).}
\begin{tabular}{lccccccccccccccc}
\hline
(1) & (2) & (3) & (4) &  (5) &  (6) & (7) &  (8) &  (9) &  (10) &    (11) &
 (12) &  (13) & (14) & (15) & (16) \\
Star & $l~(^{\circ})$ & $b~(^{\circ})$ & $V_o$ &  $(B-V)_o$ &  $(V-I)_o$ & $g_o$ &  
$(g-r)_o$ &  $(r-i)_o$ &  $J_o$ &  $(J-H)_o$ &  $(H-K_{s})_{o}$ & $W1_o$ & $(W1-W2)_o$ & $(W2-W3)_o$ & $E_{d}(B-V)$ \\
\hline
$T7934\_31569\_1$ & 2.10152  &  -28.69877 & 10.236 & 1.068 & 0.917  & 10.920 & 0.784 & 0.260 & 8.635 & 0.513 & 0.118 & 7.933 & -0.097 & 0.116 & 0.085 \\
$T5031\_00326\_1$ & 2.46213  &  +34.06951 & 15.566 & 0.804 & 0.890  & 11.296 & 0.749 & 0.247 & 9.095 & 0.503 & 0.113 & 8.478 & -0.087 & 0.135 & 0.169 \\
$T7949\_00356\_1$ & 3.90360  &  -34.07636 & 10.651 & 0.991 & 1.143  & 11.082 & 0.850 & 0.288 & 8.646 & 0.555 & 0.133 & 8.024 &  0.015 & 0.147 & 0.047 \\
         ...  &    ...   &       ...  &  ...   &   ... &    ... &  ...   &   ... &  ...  &  ...  &  ...  &  ...  &   ... &  ...   & ... & ...  \\           
         ...  &    ...   &       ...  &  ...   &   ... &    ... &  ...   &   ... &  ...  &  ...  &  ...  &  ...  &   ... &  ...   & ... & ...  \\
         ...  &    ...   &       ...  &  ...   &   ... &    ... &  ...   &   ... &  ...  &  ...  &  ...  &  ...  &   ... &  ...   & ... & ...  \\      
$T8396\_01656\_1$ & 351.88421& -33.13869  & 10.517 & 0.921 &1.165   & 11.143 & 0.874 & 0.299 & 8.647 & 0.579 & 0.130 & 7.910 & -0.118 &  0.201 & 0.041 \\
$T8409\_01526\_1$ & 352.34396& -34.52388  & 10.510 & 1.012 &1.287   & 10.864 & 0.746 & 0.247 & 8.651 & 0.544 & 0.075 & 7.939 & -0.100 &  0.158 & 0.038 \\
$T5008\_00508\_1$ & 354.27874& +45.28820  & 9.669  & 0.703 & 1.053  &  9.993 & 0.656 & 0.209 & 7.974 & 0.532 & 0.023 & 7.343 & -0.031 & -0.042 & 0.183 \\ 
\hline
\end{tabular}
}
\end{table*}
\begin{equation}
I_{c}=1.040(\pm0.007)\times I_{d}-0.501(\pm0.085).
\end{equation}
The $V$ magnitudes and $B-V$ colours were evaluated by the following equations taken from the {\em Hipparcos and Tyhco} catalogue \citep{ESA97}:

\begin{eqnarray}
V=V_{T}+0.0036-0.1284\times (B_{T}-V_{T})+0.0442\times (B_{T}-V_{T})^{2}\nonumber \\
~~~~-0.015\times (B_{T}-V_{T})^{3},\nonumber \\
(B-V)=(B_{T}-V_{T})-0.113-0.258\times z+0.40\times z^{3},\nonumber \\  
if~~~0.65<(B_{T}-V_{T})<1.1~~~and~~~z=(B_{T}-V_{T})-0.95, \\ \nonumber   
(B-V)=(B_{T}-V_{T})-0.173-0.220\times z-0.01\times z^{3},\nonumber \\
if~~~1.1<(B_{T}-V_{T})~~~and~~~z=(B_{T}-V_{T})-1.20.\nonumber
\end{eqnarray}

\subsection{Reddening and Metallicity}
The $E(B-V)$ colour excess of stars have been evaluated in two steps. First, we used the maps taken from \citet{Schlegel98} and evaluated a $E_{\infty}(B-V)$ colour excess for each star. We then reduced them using the following procedure \citep{Bahcall80}:
\begin{equation}
A_{d}(b)=A_{\infty}(b)\Biggl[1-\exp\Biggl(\frac{-\mid d \times \sin(b)\mid}{H}\Biggr)\Biggr].
\end{equation}
Here, $b$ and $d$ are the Galactic latitude and distance of the star, respectively. $H$ is the scaleheight for the interstellar dust which is adopted as 125 pc \citep{Marshall06} and $A_{\infty}(b)$ and $A_{d}(b)$ are the total absorptions for the model and for the distance to the star, respectively. $A_{\infty}(b)$ can be evaluated by means of the following equation:

\begin{equation}
A_{\infty}(b)=3.1\times E_{\infty}(B-V).
\end{equation}
$E_{\infty}(B-V)$ is the colour excess for the model taken from the \citet{Schlegel98}. Then, $E_{d}(B-V)$, i.e. the colour excess for the corresponding star at the distance $d$, can be evaluated by Eq. (5) adopted for distance $d$,

\begin{equation}
E_{d}(B-V)=A_{d}(b)~/~3.1.
\end{equation}
As explained in Section 2.1, our sample consists of RC stars. Hence, we adopted the absolute magnitude $M_{K_s}=-1.54\pm0.04$ cited by \citep{Groenewegen08} and substituted it into the following equation to obtain the distances of the sample stars:
\begin{equation}
(K_{s}-M_{K_s})_{0}=5\times \log d-5
\end{equation}
This value has also a small dependence on metallicity and age, hence it can be used accurately in determining the distances to the sources (see \citet{Cabrera07} and references therein for a complete description about using the red clump sources as distance estimators). As the total absorptions for the model and distance to a star are different, $A_{d}(b)$, and colour excess, $E_{d}(B-V)$, could be evaluated by iterating Eqs. (3)-(6).

We have omitted the indices $\infty$ and $d$ from the colour excess $E(B-V)$ in the equations. However, we use the terms model for the colour excess of \citet{Schlegel98} and ``reduced'' the colour excess corresponding to distance $d$. The total absorption $A_{d}$ used in the section and classical total absorption $A_{V}$ have the same meaning. 

We de-reddened the colours and magnitudes by using the $E_{d}(B-V)$ colour excesses of the stars evaluated using the procedures explained above and the equations of \citet{Fan99} and \citet{Fiorucci03} for $V-I$ colour and for the 2MASS photometry. 

\begin{figure*}
\begin{center}
\includegraphics[scale=0.80, angle=0]{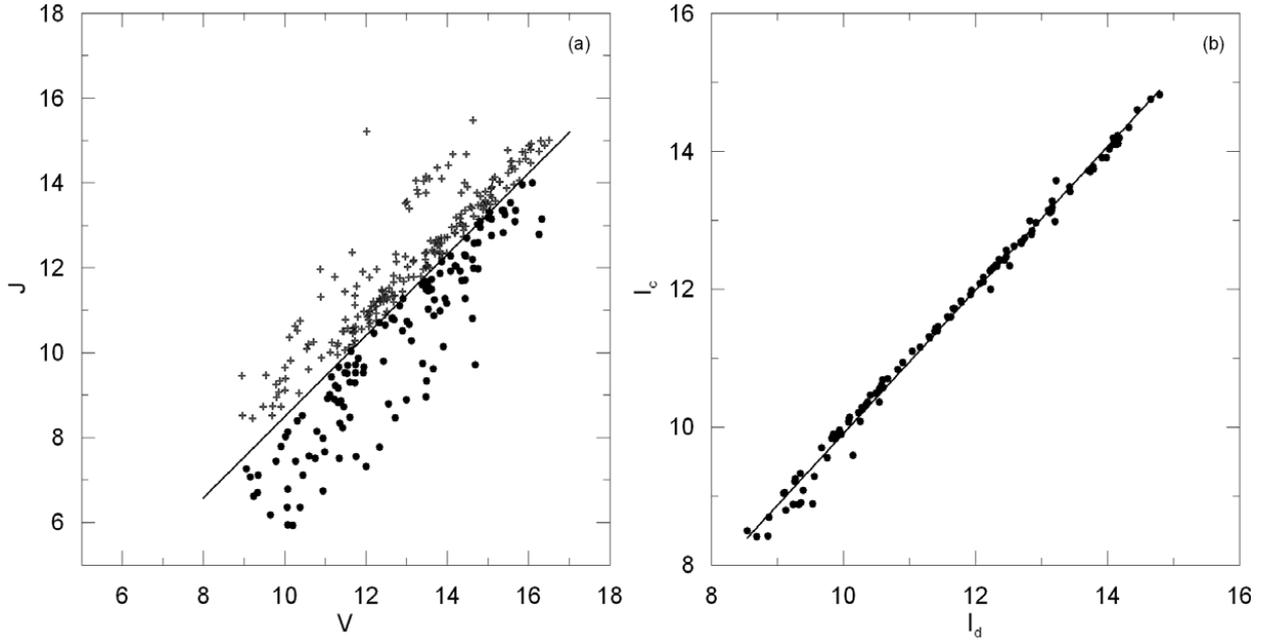}
\caption[] {The $J$ versus $V$ two magnitude diagram for 344 stars observed by \citet{Landolt09} and in 2MASS survey (panel a) and the two magnitude diagram of Johnson's and DENIS optical magnitudes, $I_{c}$ and $I_{d}$, for the 124 giants (panel b) identified in panel (a).}
\end{center}
\end{figure*}

\begin{eqnarray}
V_{o}=V-3.1\times E_d(B-V),\nonumber \\
(B-V)_{o}= (B-V)-E_d(B-V),\nonumber \\    
(V-I)_{o}= (V-I)-1.250\times E_d(B-V),\\ \nonumber 
J_{o}=J-0.887\times E_d(B-V), \nonumber \\
(J-H)_{o}= (J-H)-0.322\times E_d(B-V),\nonumber \\
(H-K_{s})_{o}=(H-K_{s})-0.183\times E_d(B-V).\nonumber
\end{eqnarray}

As the intrinsic $gri$ magnitudes were transformed from the equations of \citet{Yaz10}, no de-reddening were necessary for them. For de-reddening the magnitudes $W1$, $W2$, and $W3$, we adopted the corresponding total absorptions cited by \citet{Bilir11a} i.e. $A_{W1}/A_V=0.051$, $A_{W2}/A_V=0.030$, and $A_{W3}/A_V=0.028$, evaluated by means of a spline function fitted to the data of \citet{Cox00} which cover a range of $0.002~\mu m \leq \lambda \leq 250~\mu m$. Fig. 5 shows that the colour excess, $E(B-V)$, is less than 0.1 for most of the stars, and that their distribution peaks at $E(B-V)$=0.05 mag.

The complete data for the sample of 355 stars are given in Table 1, while the errors for the magnitudes and colours for $BVI$, $JHK_{s}$, and $W1W2W3$ photometric systems are given in Table 2 and Fig. 6. As the SDSS magnitudes were transformed from \citet{Yaz10}, we have not shown the corresponding errors in this study. The metallicities are the calibrated values of RAVE DR3. Fig. 7 shows that our sample consists mostly of thin and thick disc stars, that present mean metallicities of -0.4 dex \citep{Rocha06}, and -0.7 dex \citep{Cabrera05}, respectively. There are only very few stars with $[M/H]<-1$ and $[M/H]>0.2$ dex, and the mode is at $[M/H]=-0.35$ dex.

\begin{table}
\center 
\caption{Mean errors and standard deviations for the magnitudes of Johnson-Cousins, 2MASS and {\it WISE} photometries.}
\begin{tabular}{ccccc}
\hline
 Filter & Mean error (mag) & $s$           & Photometry \\
\hline
         B & 0.1155 &      0.0370  &Johnson-Cousins\\
         V & 0.0604 &      0.0161  &               \\
         I & 0.0349 &      0.0531  &               \\
         J & 0.0246 &      0.0044  &  2MASS        \\
         H & 0.0382 &      0.0123  &               \\
   $K_{s}$ & 0.0244 &      0.0046  &               \\
        W1 & 0.0246 &      0.0043  & {\it WISE}    \\
        W2 & 0.0216 &      0.0032  &               \\
	W3 & 0.0393 &      0.0098  &               \\
\hline
\end{tabular}
\end{table}

\begin{figure}
\begin{center}
\includegraphics[scale=0.40, angle=0]{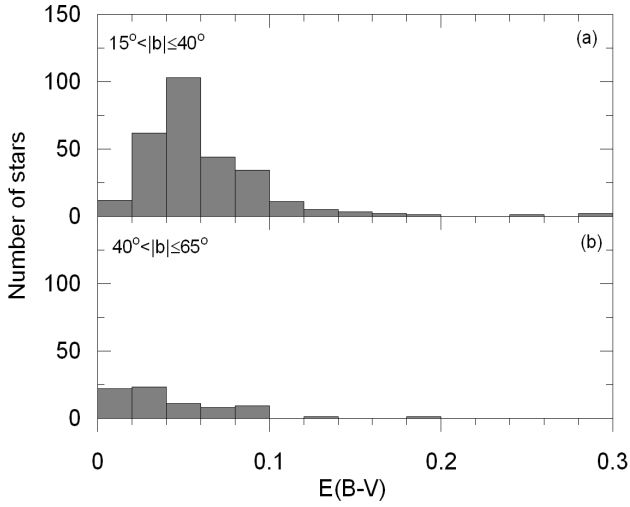}
\caption[] {Distribution of colour excess $E(B-V)$.}
\end{center}
\end{figure}

\begin{figure*}
\begin{center}
\includegraphics[scale=0.85, angle=0]{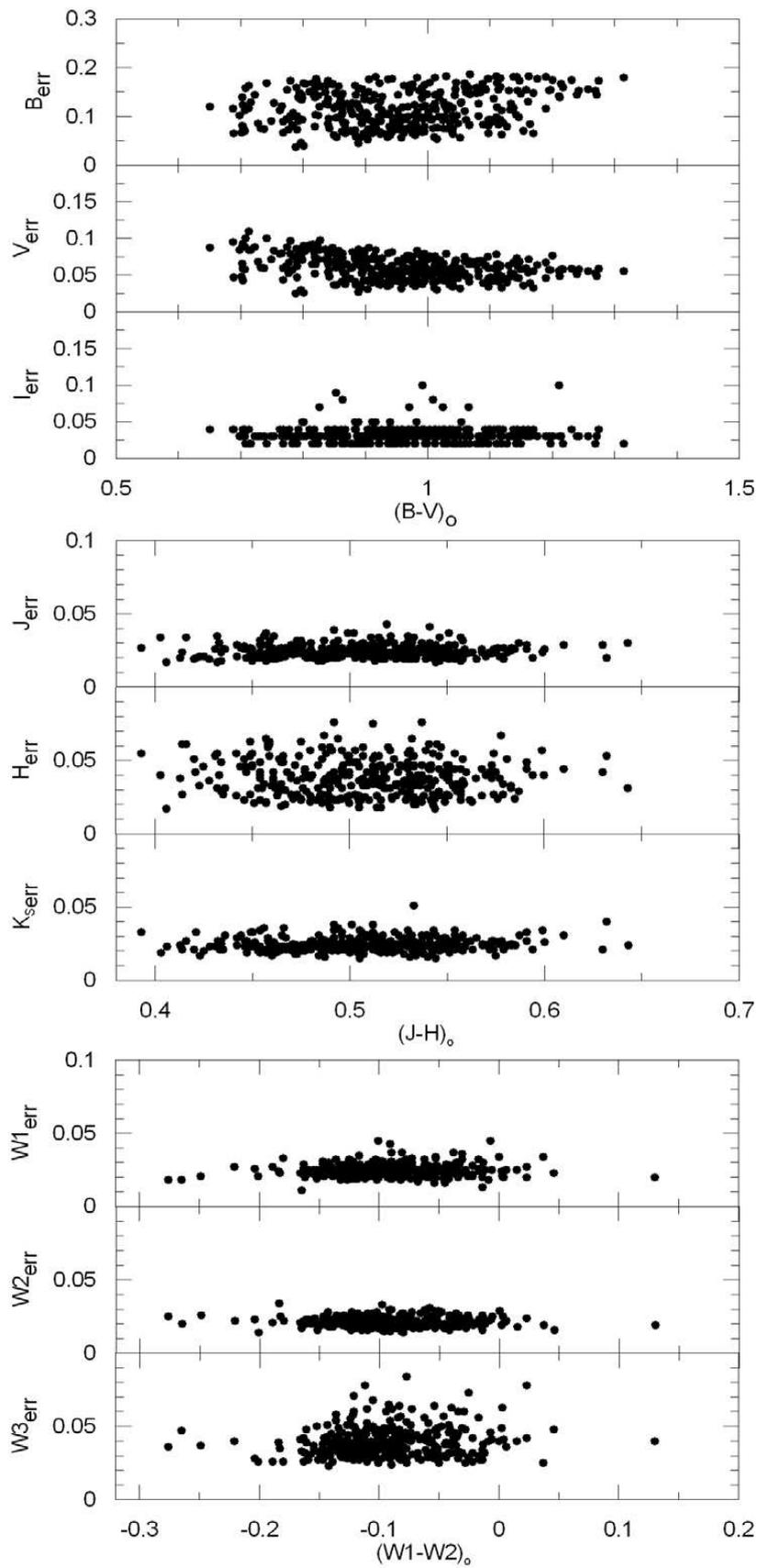}
\caption[] {The error distributions for Johnson-Cousins ($BVI$), 2MASS ($JHK_s$), and {\it WISE} ($W1W2W3$) magnitudes.}
\end{center}
\end{figure*}

\section{Results}
\subsection{Metallicity dependent transformations between {\it WISE} and Johnson-Cousins, SDSS, 2MASS photometries}

We adopted the procedure in \citet{Yaz10} and used the following general equations to derive nine sets of transformations between {\it WISE} and Johnson-Cousins, SDSS, and 2MASS photometries. As explained in \citet{Yaz10}, this approach, that includes a metallicity term instead of deriving transformations for a set of stars with a metallicity range but omitting the metallicity term, can be explained by the fact that stars change their positions in two colour diagrams by shifting an amount proportional to their metallicities. The general equations are as follows:

\begin{eqnarray}
(V-W1)_{0}=a_{1}(B-V)_{0}+b_{1}(V-I)_{0}+c_{1}[M/H]+d_{1},\nonumber \\
(V-W2)_{0}=a_{2}(B-V)_{0}+b_{2}(V-I)_{0}+c_{2}[M/H]+d_{2},\nonumber\\  
(V-W3)_{0}=a_{3}(B-V)_{0}+b_{3}(V-I)_{0}+c_{3}[M/H]+d_{3},\nonumber \\ 
(g-W1)_{0}=a_{4}(g-r)_{0}+b_{4}(r-i)_{0}+c_{4}[M/H]+d_{4},\nonumber \\
(g-W2)_{0}=a_{5}(g-r)_{0}+b_{5}(r-i)_{0}+c_{5}[M/H]+d_{5},\\ \nonumber 
(g-W3)_{0}=a_{6}(g-r)_{0}+b_{6}(r-i)_{0}+c_{6}[M/H]+d_{6},\nonumber \\
(J-W1)_{0}=a_{7}(J-H)_{0}+b_{7}(H-K_s)_{0}+c_{7}[M/H]+d_{7},\nonumber \\
(J-W2)_{0}=a_{8}(J-H)_{0}+b_{8}(H-K_s)_{0}+c_{8}[M/H]+d_{8},\nonumber \\ 
(J-W3)_{0}=a_{9}(J-H)_{0}+b_{9}(H-K_s)_{0}+c_{9}[M/H]+d_{9}.\nonumber \\ \nonumber 
\end{eqnarray}

The first three sets correspond to the Johnson-Cousins photometry, while the second three and third three sets were derived for SDSS and 2MASS photometries. The numerical values of the coefficients in Eqs. (8) are given in Table 3.

\begin{table*}
\setlength{\tabcolsep}{1.5pt}
{\small
\centering
\caption{Coefficients $a_i$, $b_i$, $c_i$, and $d_i$ for the metal dependent transformations in Eqs. (8) and the coefficients $\alpha_i$, $\beta_i$ and $\gamma_i$ for the metal-free transformations in Eqs. (9) in column matrix form. The subscript $i$=1, 2, 3, 4, 5, 6, 7, 8, and 9 correspond to the same number that denotes the columns. $R$, $s$, and $<\Delta_{res}>$ denotes the correlation coefficient, standard deviation and the mean of residual for the colour indicated at the top.}
\begin{tabular}{cccccccccc}
\hline
             &     $i=1$       &    $i=2$        &     $i=3$ 	    &     $i=4$        &     $i=5$        &     $i=6$        &     $i=7$       &     $i=8$       &     $i=9$         \\
             &  $(V-W1)_o$     &   $(V-W2)_o$    &    $(V-W3)_o$    &   $(g-W1)_o$     &   $(g-W2)_o$     &   $(g-W3)_o$     &   $(J-W1)_o$    &   $(J-W2)_o$    &  $(J-W3)_o$       \\
\hline  
       $a_i$ & 0.231$\pm$0.056 & 0.206$\pm$0.054 &  0.228$\pm$0.056 & 12.781$\pm$1.272 & 11.618$\pm$0.973 & 11.513$\pm$1.453 & 1.081$\pm$0.056 & 0.946$\pm$0.041 & 1.004$\pm$0.066   \\
       $b_i$ & 0.859$\pm$0.049 & 0.822$\pm$0.047 &  0.859$\pm$0.049 &-23.488$\pm$3.074 &-20.908$\pm$2.351 &-20.510$\pm$3.510 & 0.695$\pm$0.076 & 0.618$\pm$0.056 & 0.731$\pm$0.089   \\
       $c_i$ & 0.177$\pm$0.031 & 0.161$\pm$0.030 &  0.179$\pm$0.031 & -0.067$\pm$0.013 & -0.077$\pm$0.010 & -0.054$\pm$0.015 & 0.016$\pm$0.010 &-0.003$\pm$0.007 & 0.019$\pm$0.012   \\
       $d_i$ & 1.367$\pm$0.086 & 1.330$\pm$0.083 &  1.410$\pm$0.086 & -0.911$\pm$0.197 & -0.774$\pm$0.151 & -0.654$\pm$0.225 & 0.057$\pm$0.034 & 0.035$\pm$0.025 & 0.133$\pm$0.039   \\
         $R$ & 0.758           & 0.754	         &  0.759           &  0.974           &  0.983           &  0.965           & 0.749           & 0.795           & 0.678             \\
         $s$ & 0.127           & 0.121           &  0.127           &  0.045           &  0.035           &  0.052           & 0.043           & 0.031           & 0.050             \\
$<\Delta_{res}>$ &-0.00052       & 0.00014  	        & -0.00017         &  0.00002         &  0.00062         &  0.00084         &-0.00009         & -0.00016        &-0.00022    \\

\hline
             &     $i=1$       &    $i=2$        &     $i=3$ 	    &     $i=4$        &     $i=5$        &     $i=6$        &     $i=7$       &     $i=8$       &     $i=9$         \\
             &  $(V-W1)_o$     &   $(V-W2)_o$    &    $(V-W3)_o$    &   $(g-W1)_o$     &   $(g-W2)_o$     &   $(g-W3)_o$     &   $(J-W1)_o$    &   $(J-W2)_o$    &  $(J-W3)_o$       \\ 
\hline
       $\alpha_i$ & 0.362$\pm$0.053 & 0.326$\pm$0.051 &  0.362$\pm$0.053 &  9.301$\pm$1.126 &  7.620$\pm$0.900 &  8.685$\pm$1.263 & 1.102$\pm$0.055 & 0.942$\pm$0.040 & 1.029$\pm$0.064   \\
       $\beta_i$ & 0.941$\pm$0.049 & 0.897$\pm$0.047 &  0.943$\pm$0.049 &-15.287$\pm$2.745 &-11.485$\pm$2.193 &-13.843$\pm$3.078 & 0.737$\pm$0.072 & 0.610$\pm$0.053 & 0.780$\pm$0.084   \\
       $\gamma_i$ & 1.082$\pm$0.073 & 1.071$\pm$0.070 &  1.121$\pm$0.073 & -0.306$\pm$0.166 & -0.079$\pm$0.133 & -0.162$\pm$0.186 & 0.035$\pm$0.031 & 0.039$\pm$0.022 & 0.106$\pm$0.036   \\
         $R$ & 0.731           & 0.729           &  0.731           &  0.972           &  0.980           &  0.963           & 0.746           & 0.794           & 0.675             \\
         $s$ & 0.132           & 0.126           &  0.132           &  0.047           &  0.038           &  0.053           & 0.043           & 0.031           & 0.050             \\
$<\Delta_{res}>$ & 0.00128         & 0.00030         & -0.00011         & -0.00010         &  0.00024         &-0.00042          & -0.00031        & 0.00010         & 0.00048          \\

\hline
\end{tabular} 
} 
\end{table*}

\begin{figure}
\begin{center}
\includegraphics[scale=0.45, angle=0]{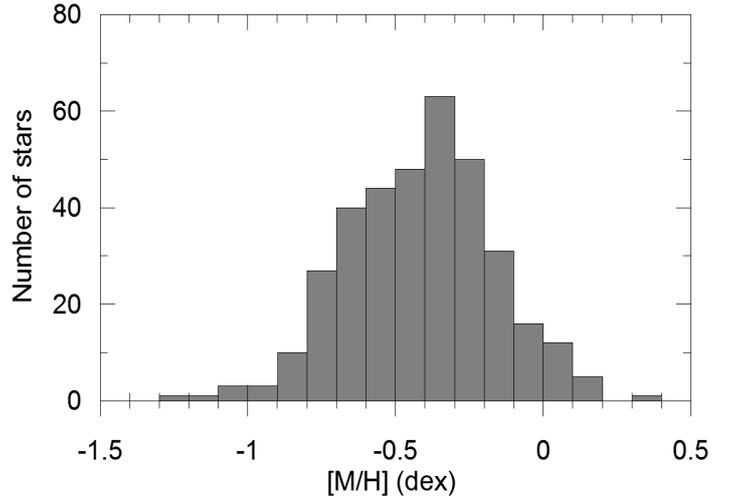}
\caption[] {Metallicity distribution for the sample stars.}
\end{center}
\end{figure}

\subsection{Metal-free transformations between {\it WISE} and Johnson-Cousins, SDSS, 2MASS photometries}
 
We derived also metal-free transformations between {\it WISE} and Johnson-Cousins, SDSS, and 2MASS photometries. These can be used to transform the $BVI$, $gri$, and $JHK_s$ data of RC stars with unknown metallicities. Thus we give the chance to the researchers to transfer their $BVI$, $gri$, and $JHK_{s}$ data with lack of metallicities for RC stars to the {W1W2W3} ones. The general equations are as follows: 
 
\begin{eqnarray}
(V-W1)_{0}=\alpha_{1}(B-V)_{0}+\beta_{1}(V-I)_{0}+\gamma_{1}, \nonumber \\ 
(V-W2)_{0}=\alpha_{2}(B-V)_{0}+\beta_{2}(V-I)_{0}+\gamma_{2},\nonumber \\
(V-W3)_{0}=\alpha_{3}(B-V)_{0}+\beta_{3}(V-I)_{0}+\gamma_{3},\nonumber \\
(g-W1)_{0}=\alpha_{4}(g-r)_{0}+\beta_{4}(r-i)_{0}+\gamma_{4},\nonumber \\
(g-W2)_{0}=\alpha_{5}(g-r)_{0}+\beta_{5}(r-i)_{0}+\gamma_{5},\\ \nonumber
(g-W3)_{0}=\alpha_{6}(g-r)_{0}+\beta_{6}(r-i)_{0}+\gamma_{6},\nonumber \\
(J-W1)_{0}=\alpha_{7}(J-H)_{0}+\beta_{7}(H-K_s)_{0}+\gamma_{7},\nonumber \\
(J-W2)_{0}=\alpha_{8}(J-H)_{0}+\beta_{8}(H-K_s)_{0}+\gamma_{8},\nonumber \\
(J-W3)_{0}=\alpha_{9}(J-H)_{0}+\beta_{9}(H-K_s)_{0}+\gamma_{9}.\nonumber\\ \nonumber 
\end{eqnarray}
    
The numerical values of the coefficients in Eqs. (9) are given in Table 3. The comparison between the correlation coefficients and the standard deviations for Eqs. (8) and (9) show that metallicity dependent transformations are the preferred ones.

\subsection{Metallicity dependent inverse transformations between {\it WISE} and Johnson-Cousins, SDSS photometries}
As was explained before, $W4$ magnitudes cannot be used for the RC star sample. Hence, we adapted the procedure used for dwarfs \citep{Bilir11a} to get the metallicity dependent inverse transformations with two colours: By combining linearly the near and mid-infrared colours,  $(J-H)_{o}$ and $(W1-W2)_{o}$, we transformed them to the optical colours: $(B-V)_{o}$ and $(V-I)_{o}$, $(g-r)_{o}$ and $(r-i)_{o}$. The general equations are as follows: 

\begin{eqnarray}
(B-W1)_{0}=a_{1}(J-H)_{0}+b_{1}(W1-W2)_{0}+c_{1}[M/H]+d_{1},\nonumber \\
(V-W1)_{0}=a_{2}(J-H)_{0}+b_{2}(W1-W2)_{0}+c_{2}[M/H]+d_{2},\nonumber \\
(I-W1)_{0}=a_{3}(J-H)_{0}+b_{3}(W1-W2)_{0}+c_{3}[M/H]+d_{3},\nonumber \\
(g-W1)_{0}=a_{4}(J-H)_{0}+b_{4}(W1-W2)_{0}+c_{4}[M/H]+d_{4},\\ \nonumber
(r-W1)_{0}=a_{5}(J-H)_{0}+b_{5}(W1-W2)_{0}+c_{5}[M/H]+d_{5},\nonumber \\
(i-W1)_{0}=a_{6}(J-H)_{0}+b_{6}(W1-W2)_{0}+c_{6}[M/H]+d_{6}. \nonumber \\ \nonumber 
\end{eqnarray}
The numerical values of the coefficients in Eqs. (10) are given in Table 4.                         

\textwidth=23.1cm
\begin{landscape}
\begin{figure*}
\includegraphics[scale=1.1, angle=0]{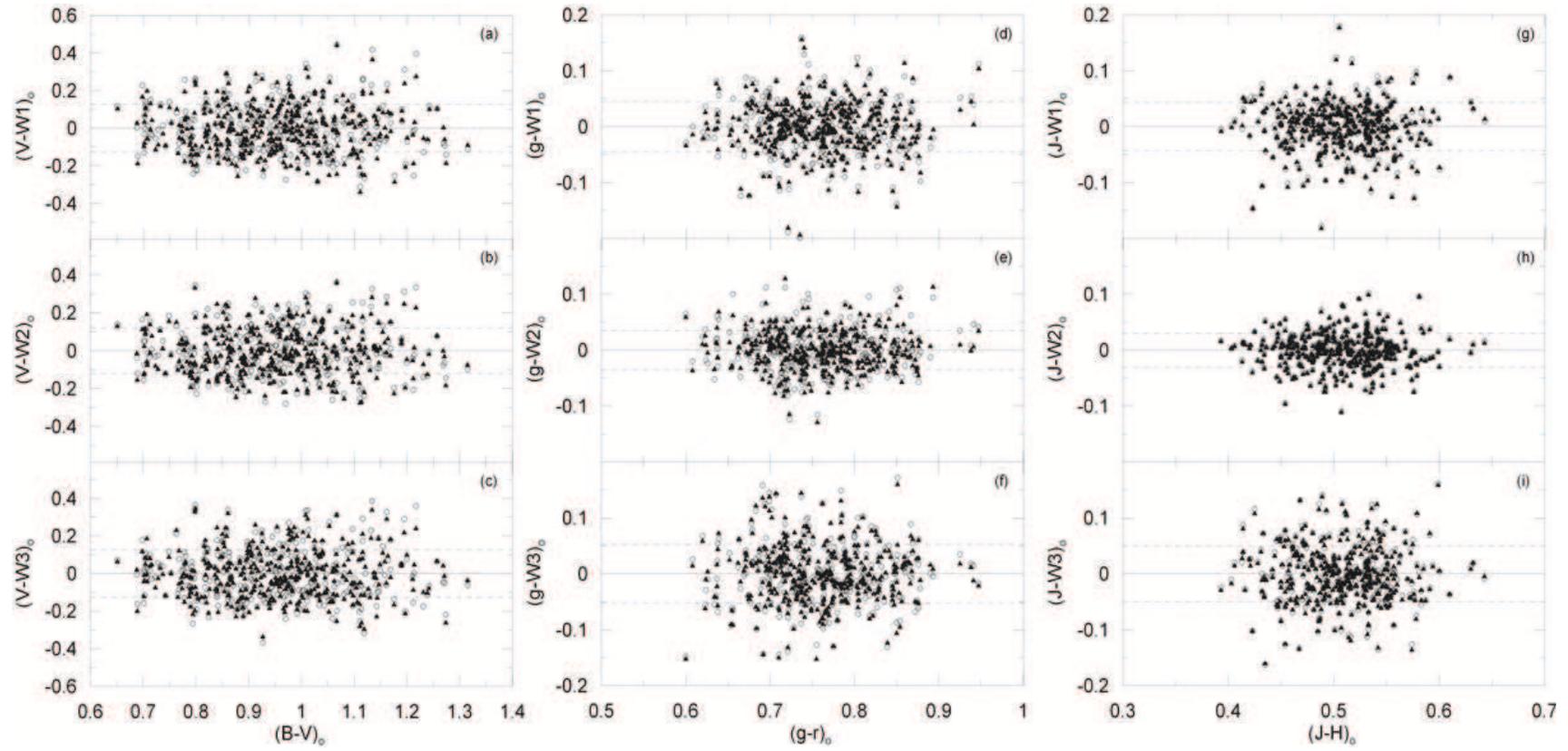}
\caption[] {Colour residuals for the metallicity dependent ($o$) and metal free ($+$) transformations. The notation used is $\Delta$(colour)=(evaluated colour)-(observed colour). The horizontal dashed lines correspond to 1$\sigma$ residuals.}
\end{figure*}
\end{landscape}

\textwidth=18.1cm
\begin{table*}
\setlength{\tabcolsep}{4pt}
\centering
\caption{Coefficients $a_i$, $b_i$,$c_i$, and $d_i$ for the metal dependent inverse transformations in Eqs. (10) and the coefficients $\alpha_i$, $\beta_i$ and $\gamma_i$ for the metal-free inverse transformations in Eqs. (11) in column matrix form. The subscript $i$=1, 2, 3, 4, 5, and 6 correspond to the same number that denotes the the columns. $R$, $s$, and $<\Delta_{res}>$ denotes the correlation coefficient, standard deviation and the mean of residual for the colour indicated at the top.}
\begin{tabular}{cccccccccc}
\hline
             &     $i=1$       &    $i=2$        &     $i=3$ 	 &        $i=4$        &     $i=5$        &     $i=6$         \\
             &  $(B-W1)_o$     &  $(V-W1)_o$     &  $(I-W1)_o$ 	 &   $(g-W1)_o$        &  $(r-W1)_o$      &  $(i-W1)_o$       \\
\hline
       $a_i$ & 2.980$\pm$0.212 & 2.380$\pm$0.167 &  1.480$\pm$1.140 &  3.352$\pm$0.124 &  2.350$\pm$0.080 &  1.945$\pm$0.063  \\
       $b_i$ &-0.911$\pm$0.194 &-0.845$\pm$0.153 & -0.633$\pm$0.128 & -0.804$\pm$0.113 & -0.764$\pm$0.073 & -0.747$\pm$0.058  \\
       $c_i$ & 0.452$\pm$0.037 & 0.264$\pm$0.029 &  0.146$\pm$0.024 &  0.223$\pm$0.022 &  0.136$\pm$0.014 &  0.102$\pm$0.011  \\
       $d_i$ & 1.940$\pm$0.111 & 1.220$\pm$0.088 &  0.620$\pm$0.073 &  1.224$\pm$0.065 &  0.941$\pm$0.042 &  0.882$\pm$0.033  \\
         $R$ & 0.758           & 0.732           &  0.624           &  0.870           &  0.888           &  0.899            \\
	 $s$ & 0.168           & 0.132           &  0.110           &  0.098           &  0.064           &  0.050            \\
$<\Delta_{res}>$ & -0.00070    &-0.00109         & -0.00121         &  0.00057         & -0.00012         & -0.00019          \\

\hline
             &     $i=1$       &     $i=2$       &     $i=3$        &     $i=4$        &     $i=5$        &     $i=6$         \\
             &  $(B-W1)_o$     &  $(V-W1)_o$     &  $(I-W1)_o$ 	    &   $(g-W1)_o$     &  $(r-W1)_o$      &  $(i-W1)_o$       \\
\hline
       $\alpha_i$ & 3.353$\pm$0.251 & 2.597$\pm$0.184 &  1.598$\pm$0.145 &  3.536$\pm$0.140 &  2.462$\pm$0.090 &  2.129$\pm$0.070 \\
       $\beta_i$  &-1.199$\pm$0.230 &-1.014$\pm$0.168 & -0.726$\pm$0.133 & -0.947$\pm$0.128 & -0.851$\pm$0.082 & -0.812$\pm$0.064  \\
       $\gamma_i$ & 1.534$\pm$0.127 & 0.983$\pm$0.093 &  0.489$\pm$0.073 &  1.025$\pm$0.071 &  0.819$\pm$0.045 &  0.791$\pm$0.035 \\
         $R$ & 0.627           & 0.653           &  0.570           &  0.825           &  0.854           &  0.872                \\
         $s$ & 0.200           & 0.147           &  0.116           &  0.112           &  0.072           &  0.056                \\
$<\Delta_{res}>$ & 0.00000     &-0.00048         &  0.00016         & -0.00046         &  0.00005         & -0.00055              \\
\hline
\end{tabular}  
\end{table*} 

\subsection{Metal-free inverse transformations between {\it WISE} and Johnson-Cousins, SDSS photometries}
We adapted the procedure explained in Section 3.3 and derived metal-free inverse transformations between {\it WISE} and Johnson-Cousins, SDSS photometries. The general equations are as follows, and the numerical values of the coefficients in these equations are given in Table 4:

\begin{eqnarray}
(B-W1)_{0}=\alpha_{1}(J-H)_{0}+\beta_{1}(W1-W2)_{0}+\gamma_{1}, \nonumber \\ 
(V-W1)_{0}=\alpha_{2}(J-H)_{0}+\beta_{2}(W1-W2)_{0}+\gamma_{2}, \nonumber \\ 
(I-W1)_{0}=\alpha_{3}(J-H)_{0}+\beta_{3}(W1-W2)_{0}+\gamma_{3}, \nonumber \\ 
(g-W1)_{0}=\alpha_{4}(J-H)_{0}+\beta_{4}(W1-W2)_{0}+\gamma_{4}, \\ \nonumber
(r-W1)_{0}=\alpha_{5}(J-H)_{0}+\beta_{5}(W1-W2)_{0}+\gamma_{5}, \nonumber \\ 
(i-W1)_{0}=\alpha_{6}(J-H)_{0}+\beta_{6}(W1-W2)_{0}+\gamma_{6}. \nonumber \\ \nonumber
\end{eqnarray}
Comparison of the correlation coefficients and the standard deviations for Eqs. (10) and (11) show that the inverse transformations are recommended especially when they are used with a metallicity term as in the direct transformations.

\subsection{Residuals}
We compared the observed colours with those evaluated by means of the transformations. The residuals corresponding to the Eqs. (8) and (9) are plotted versus observed colours $(B-V)_0$, $(g-r)_0$, and $(J-H)_0$ in the same figure (Fig. 8) with different symbols. For the observed optical colours, the residuals corresponding to the equations just cited are different and they favour the ones with metallicity term. Whereas for the observed near-infrared colour, i.e. $(J-H)_0$, the two sets of residuals overlap, diminishing the effect of the metallicity term. The same result can be deduced from comparison of the metallicity term $c_i$ (i=1,...,9) in Table 3, where $c_i$ decreases from $0.179\pm0.031$ for $(V-W3)_0$ to $-0.003\pm0.007$ for $(J-W2)_0$. The residuals corresponding to the Eqs. (10) and (11) are plotted versus observed $(W1-W2)_0$ in the same figure with different symbols (Fig. 9). The difference between the  residuals of two sets are much larger for the ones corresponding to the $BVI$ magnitudes relative to the residuals for $gri$. The numerical values of the metallicity term, $c_i$ (i =1,...,6), in Table 4 confirm this suggestion. Actually, $c_1=0.452\pm0.037$ for $(B-W1)_0$ whereas it is only $c_6=0. 102\pm0.011$ for $(i-W1)_0$. Hence, we conclude that the metallicity term provides more accurate inverse transformations for $BVI$ magnitudes, but that its contribution to $gri$ is rather limited.

\begin{figure*}
\begin{center}
\includegraphics[scale=0.85, angle=0]{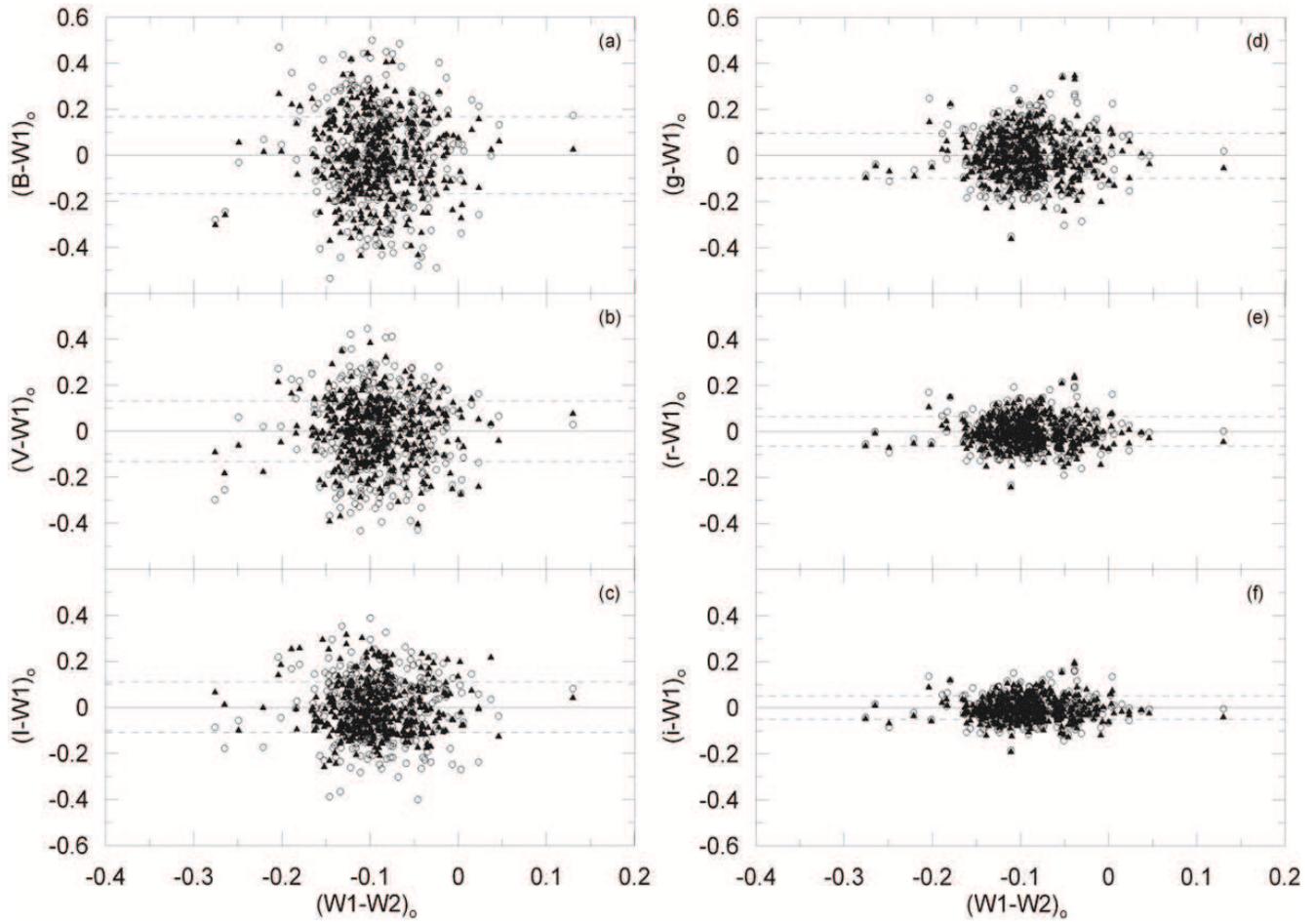}
\caption[] {Colour residuals for the metallicity dependent ($o$) and metal free ($+$) inverse transformations. The notation used is $\Delta$(colour)=(evaluated colour)-(observed colour). The horizontal dashed lines correspond to 1$\sigma$ residuals.}
\end{center}
\end{figure*}

\section{Summary and conclusion}
We have obtained colour transformations for the conversion of {\it WISE} ($W1W2W3$) magnitudes to the Johnson-Cousins' ($BVI$), SDSS ($gri$), and 2MASS ($JHK_s$) photometric systems, for RC stars. The sample were selected by applying two constraints: 1) $2<\log g~(cms^{-2})<3$, and 2) $J-H>0.4$, to the RAVE DR3 data (resulting a sample of 8003 giants), and matching the coordinates of this sample with the {\em Tycho-2}, DENIS, 2MASS, and {\it WISE} catalogues we produced a reduced sample with available magnitudes that is the one used in the transformations. In order to obtain the most accurate transformations, we included four additional constraints: 3) the data were de-reddened, 4) only the stars with high quality were selected, 5) a metallicity term was added to the transformation equations, and 6) transformation equations are two-colour dependent; that reduced the total sample to 355 stars.

The transformation equations, and the inverse ones, were designed in two sheets: one with a metallicity term and the other with metallicity-free. Comparison between the correlation coefficients and the standard deviations for the two sets promotes the use of the metallicity dependent transformation equations. It is noticeable that even when the procedure used for the transformations for dwarfs was different in \citet{Bilir08, Bilir11a}, we separated the dwarf sample into different metallicity sub-samples instead of adding a metallicity term to the transformation equations, we obtained here the same result, that is they were metallicity dependent. This dependence of the transformations on metallicity had been also confirmed in \citet{Yaz10}.

As in the case of dwarfs, {\it WISE} has an advantage relative to the 2MASS photometric system due to its deeper magnitudes. Actually, $W1$ is a magnitude deeper than $K_{s}$ for sources with spectra close to an A0 star and even deeper for $K$ and $M$ spectral stars. The present transformations can be applied to stars with known absolute $V$, $g$, or $J$ magnitudes, when absolute magnitudes for $W1$ can be also provided. These two advantages can be used to investigate the RC stars in the thin and thick discs more accurately, and combining this study with the one carried out for dwarfs would be even more fruitful. A possible interesting application of the transformations presented here would be the comparison of the (new) Galactic model parameters and the ones estimated in situ, but the transformations also can be used in a wide variety of research fields.

\section{Acknowledgments}

This work has been supported in part by the Scientific and
Technological Research Council (T\"UB\.ITAK) 111T650.

Funding for RAVE has been provided by: the Australian Astronomical Observatory; 
the Leibniz-Institut fuer Astrophysik Potsdam (AIP); the Australian National 
University; the Australian Research Council; the French National Research 
Agency; the German Research Foundation; the European Research Council 
(ERC-StG 240271 Galactica); the Istituto Nazionale di Astrofisica at Padova; 
The Johns Hopkins University; the National Science Foundation of the USA 
(AST-0908326); the W. M. Keck foundation; the Macquarie University; the 
Netherlands Research School for Astronomy; the Natural Sciences and 
Engineering Research Council of Canada; the Slovenian Research Agency; the 
Swiss National Science Foundation; the Science \& Technology Facilities 
Council of the UK; Opticon; Strasbourg Observatory; and the Universities of 
Groningen, Heidelberg and Sydney. The RAVE web site is at 
http://www.rave-survey.org

This research has made use of the NASA/IPAC Infrared Science Archive and 
Extragalactic Database (NED) which are operated by the Jet Propulsion Laboratory, 
California Institute of Technology, under contract with the National Aeronautics 
and Space Administration.

This publication makes use of data products from the Two Micron All
Sky Survey, which is a joint project of the University of
Massachusetts and the Infrared Processing and Analysis
Center/California Institute of Technology, funded by the National
Aeronautics and Space Administration and the National Science
Foundation.

This research has made use of the SIMBAD, and NASA\rq s Astrophysics 
Data System Bibliographic Services.

\end{document}